\documentclass[aps,prc,twocolumn,superscriptaddress]{revtex4-1}

\usepackage{graphicx}
\usepackage{amsmath}
\usepackage{multirow}
\usepackage{bm}
\usepackage[colorlinks,
linkcolor=blue,
anchorcolor=blue,
urlcolor=blue,
citecolor=blue]{hyperref}

\begin{document}

\title{ Resonances of $A=4$ $T=1$ isospin triplet states within the \textit{ab initio} no-core Gamow shell model}


\author{J. G. Li}
\affiliation{School of Physics,  and   State Key  Laboratory  of  Nuclear  Physics   and  Technology, Peking University, Beijing  100871, China}
\author{N. Michel}
\affiliation{Institute of Modern Physics, Chinese Academy of Sciences, Lanzhou 730000, China}
\affiliation{School of Nuclear Science and Technology, University of Chinese Academy of Sciences, Beijing 100049, China}
\author{W. Zuo}
\affiliation{Institute of Modern Physics, Chinese Academy of Sciences, Lanzhou 730000, China}
\affiliation{School of Nuclear Science and Technology, University of Chinese Academy of Sciences, Beijing 100049, China}
\author{F. R. Xu}\email[]{frxu@pku.edu.cn}
\affiliation{School of Physics,  and   State Key  Laboratory  of  Nuclear  Physics   and  Technology, Peking University, Beijing  100871, China}



\date{\today}

\begin{abstract}
 
The $A=4$ nuclei, i.e., $^4$H, $^4$He and $^4$Li, establish an interesting isospin $T=1$ isobaric system. $^4$H and $^4$Li are unbound broad resonances, whereas $^4$He is deeply bound in its ground state but unbound in all its excited states. The present situation is that experiments so far have not given consistent data on the resonances.
Few-body calculations have well studied the scatterings of the $4N$ systems. 
In the present work, we provide many-body calculations of the broad resonance structures, in an \textit{ab initio} framework with modern realistic interactions.
It occurs that, indeed,  $^4$H, $^4$Li and excited $^4$He  are broad resonances, which is in accordance with experimental observations. 
The calculations also show that the first $1^-$ excited state almost degenerates with the $2^-$ ground state in the pair of mirror isobars of $^4$H and $^4$Li, which may suggest that the experimental data on energy and width  are the mixture of the ground state and the first excited state.
The $T = 1$ isospin triplet formed with an excited state of $^4$He and ground states of $^4$H and $^4$Li is studied, focusing on the effect of isospin symmetry breaking.

\end{abstract}


\maketitle

\section{Introduction}
Four-nucleon ($4N$) systems are the lightest nuclear systems to exhibit resonances at low energies \cite{PhysRevLett.15.300,PhysRev.161.1050,COHEN1965242,PhysRev.177.1455,MEYER1979335,SENNHAUSER1981409,FRANKE1985351,PhysRevC.42.448,PhysRevC.42.550,TILLEY19921,PhysRevC.33.2204,BELOZYOROV1986352,1987PZETF45196G,AmelinJEPT,PhysRevC.44.325,1995JETP,SIDORCHUK200454,2005EPJA24231G,PhysRevLett.57.543,PhysRevLett.72.3476}. 
With the highest isospin quantum number of $|T_\text{z}|=T=2$, the possible tetraneutron resonance has attracted considerable attention due to its pure neutron character. The recent experiment \cite{PhysRevLett.116.052501} has moved forward significantly on this matter, though the experiment, due to large experimental uncertainty, still cannot definitely answer whether the tetraneutron resonance  really exists. By reducing the isospin quantum number, $^4$H, $^4$He and $^4$Li form a unique $T=1$ isospin triplet of $4N$ resonances with many experimental data available \cite{PhysRevLett.15.300,PhysRev.161.1050,COHEN1965242,PhysRev.177.1455,MEYER1979335,SENNHAUSER1981409,FRANKE1985351,PhysRevC.42.448,PhysRevC.42.550,TILLEY19921,PhysRevC.33.2204,BELOZYOROV1986352,1987PZETF45196G,AmelinJEPT,PhysRevC.44.325,1995JETP,SIDORCHUK200454,2005EPJA24231G,PhysRevLett.57.543,PhysRevLett.72.3476,PhysRevLett.116.052501}. However, the data providing the energies and widths of the $ A=4$ resonances are neither conclusive nor even consistent with each other \cite{COHEN1965242,PhysRev.177.1455,MEYER1979335,SENNHAUSER1981409,FRANKE1985351,PhysRevC.33.2204,BELOZYOROV1986352,1987PZETF45196G,AmelinJEPT,PhysRevC.44.325,1995JETP,SIDORCHUK200454,2005EPJA24231G,PhysRevC.42.448,PhysRevC.42.550,TILLEY19921}. In the experiments where the missing mass method is used, the resonance positions in energy distributions depend on the resonance parameters \cite{Gurov2009}. Theoretical calculations are thus useful as they can provide suggestions for the parameters and constrain experimental results so as to derive the demanded resonance energies.

In theoretical studies, a relatively small number of nucleons allows theories to probe the underlying dynamics directly with enough accuracy and, therefore, to verify nuclear models as well as interactions  \cite{RevModPhys.70.743}. As commented in \cite{10.3389/fphy.2019.00251}, $3N$ systems remain relatively simple due to the absence of resonance structures in the continuum, while $4N$ systems present several resonance states and therefore provide better laboratories for the underlying theoretical tests. The isospin symmetry in $A = 4$ $T = 1$ multiplet states is expected to be broken, which is due in particular to the Coulomb force \cite{PhysRevLett.15.300,TILLEY19921,WERNTZ1963113}. For the pair of isospin mirror isobars, both proton-proton and Coulomb forces are absent in $^4$H, while $^4$Li has no neutron-neutron interaction. Therefore, the $A=4$ nuclei provide a unique isospin multiplet for the study of the isospin symmetry breaking, and also serve as a good testing ground to assess many-body correlations, continuum coupling and the influence of the Coulomb force in dripline regions.

Furthermore, for $4N$ resonances, both few-body and many-body methods are feasible. Resonance states provide richer structure information than bound states. However, the resonance is much difficult to be described due to the complex-energy characteristics, and continuum channels must be considered. Indeed, the nuclear complexity really starts at $A=4$ \cite{PhysRevC.42.550,10.3389/fphy.2019.00251}, and $A=4$ nuclei are of special interest \cite{PhysRevLett.85.944}. 
The scatterings of the $4N$ systems have been well studied by the Faddeev-Yakubovsky (FY) \cite{PhysRevC.86.044002,PhysRevC.93.044004,10.3389/fphy.2019.00251,Lazauskas2019,PhysRevC.58.58,CIESIELSKI1999199,PhysRevC.70.044002,PhysRevC.71.034004,PhysRevC.79.054007}, Alt-Grassberger-Sandhas (AGS) \cite{PhysRevC.75.014005,PhysRevLett.98.162502,PhysRevC.76.021001,DELTUVA2008471,PhysRevC.81.054002,PhysRevC.84.054010} and hyperspherical harmonics (HH) calculations \cite{Kievsky_2008,HH_Few,refId0,PhysRevC.84.054010}. 
However, the theoretical calculations of the energies and decay widths of the $A = 4$ resonance systems are still missing. In fact, besides mentioned reaction cross sections, only the FY calculation has assessed the $^4$H resonance width using an indirect method of extrapolation \cite{Lazauskas2019}.  Thus, other direct theoretical calculations (particularly for the resonance width) should be useful to understand the $A=4$ resonances.

Currently, there are a number of reliable many-body methods for the \textit{ab initio} description of nuclear states  \cite{LEIDEMANN2013158}. Prominent frameworks consist of the Green’s function Monte Carlo (GFMC) \cite{RevModPhys.87.1067}, no-core shell model (NCSM) \cite{BARRETT2013131}, coupled cluster \cite{hagen2016emergent} and  in-medium similarity renormalization group \cite{HERGERT2016165}. However, current \textit{ab initio} calculations are performed mainly in real-energy space without considering the resonance and continuum. 
The no-core Gamow shell model (NCGSM) \cite{PhysRevC.88.044318} within the complex-energy Berggren ensemble \cite{BERGGREN1968265} is an extension of the NCSM for open quantum systems, and has been applied to the $^5$He broad resonance \cite{PhysRevC.88.044318} and multi-neutron systems \cite{PhysRevLett.119.032501,PhysRevC.100.054313}.  In the present work, we will provide many-body \textit{ab initio} structure calculations of the $A=4$ broad resonances of $^4$H, $^4$Li and $^4$He. To probe the isospin symmetry breaking effect in the $T = 1$ isospin triplet of the resonances is another motivation of the present paper. Our calculations may provide further understandings of the existing data on the broad resonances.

\section{Method} 
The Berggren basis comprises bound, resonance and scattering single-particle (s.p.)~states \cite{BERGGREN1968265}.
The obtained Hamiltonian matrix in this representation is non-Hermitian but complex symmetric, and its eigenenergies are complex.
Energy and width are then provided by the real and imaginary parts of the Hamiltonian eigenvalues, respectively \cite{BERGGREN1968265}.

The completeness relation of the Berggren basis reads for a given partial wave:
\begin{eqnarray}\label{equation1}
   \sum_{n} |\phi_{n}\rangle\langle \phi_{n}| + \int_{L^{+}}|\phi(k)\rangle\langle \phi(k)| dk = 1,
\end{eqnarray}
where $| \phi(k) \rangle$ are the scattering states belonging to a $L^+$ contour of complex momenta, and $| \phi_{n} \rangle$ are the bound and  resonance states situated between the real $k$ axis and the $L^+$ contour.
In practical calculations, the integral in Eq. (\ref{equation1}) is discretized utilizing the Gauss-Legendre rule \cite{PhysRevLett.89.042501,Michel_2008,PhysRevC.83.034325}.

Many-body eigenstates are built from a linear combination of the Slater determinants $|SD_n\rangle = | u_1,...,u_A\rangle$, where $| u_k \rangle$, with $k \in 1, \dots, A$, is a s.p.~state of the Berggren basis, of bound, resonance or scattering character.
Coupling to the continuum is present at basis level, whereas many-body internucleon correlations occur via configuration mixing in the NCGSM framework \cite{PhysRevLett.89.042501,PhysRevLett.89.042502,Michel_2008}.
Note that the width of a resonance eigenstate obtained in NCGSM takes into account all possible particle-emission channels. 

The used Hamiltonian reads:
\begin{eqnarray}\label{equation2}
   H = \frac{1}{A}\sum_i^A\frac{(\bm{p}_i-\bm{p}_j)^2}{2m}+\sum_{i<j}^A V_{NN}^{i<j},
\end{eqnarray}
where $V_{NN}^{i<j}$ is a realistic nucleon-nucleon interaction.
There is no restriction on the type of interaction in the NCGSM calculation, contrary to the GFMC approach for example, where only local potentials can be used \cite{RevModPhys.87.1067}.
One can then use a local interaction, such as the Argonne $v_{18}$ potential \cite{PhysRevC.51.38}, or a non-local interaction, such as CD-Bonn \cite{PhysRevC.63.024001} or chiral interaction \cite{PhysRevC.68.041001} in the NCGSM.
Details about the numerical calculations of the two-body matrix elements in the NCGSM can be found in \cite{Michel_2008,PhysRevC.83.034325}.
The overlap method is utilized in order to identify the low-lying physical resonance states out of the numerous many-body eigenstates (see \cite{Michel_2008} for details).
To improve the convergence of calculations, the similarity renormalization group (SRG) \cite{PhysRevC.75.061001} is employed to soften the interaction. 

Though our Hamiltonian (\ref{equation2}) is intrinsic, the center of mass (CoM) degree of freedom is not removed in the NCGSM wave functions. In the Berggren basis, one cannot exactly factorize the many-body wave functions into the CoM and relative parts. However, as one is interested in the lowest energy states of fixed quantum numbers, their eigenenergies converge to the exact energies according to the generalized variational principle if the model space is sufficiently large, and that even in the absence of an exact treatment of the CoM motion \cite{PhysRevC.88.044318,PhysRevC.100.054313}. In \cite{PhysRevC.88.044318}, it was assessed that the CoM effect on the $^3$H energy is only 7 keV in the NCGSM calculation with the  N$^3$LO interaction. In the present calculations, we eventually use complex-momentum natural orbitals which are generated with the Berggren basis. Each of the natural orbitals is a combination of many different Berggren basis states, which expedites the convergence of the calculations. Therefore, our model space is large enough.

\begin{figure*}[!htb]
\includegraphics[width=2.0\columnwidth]{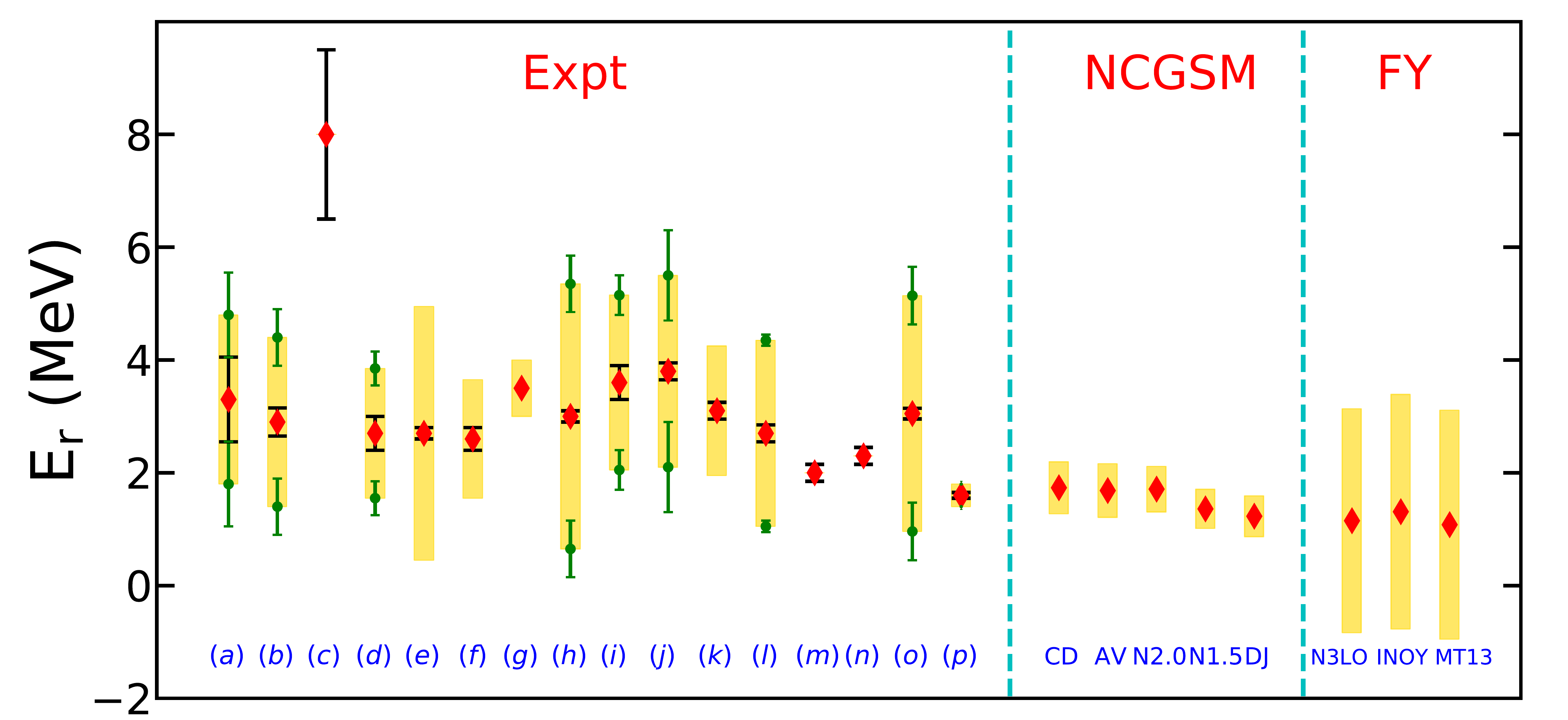}
\caption{Energy (red diamond) and width (yellow shadow) of the $2^-$ ground state of $^4$H calculated with NCGSM (central) using CD-Bonn-SRG2.0 (CD), AV18-SRG2.0 (AV), N$^3$LO-SRG2.0 (N2.0), N$^3$LO-SRG1.5 (N1.5) and Daejeon16 (DJ) (see text for definitions).
 Associated data with error bars (left) are denoted by (a-e), (f-n) and (o,p) from \cite{COHEN1965242,PhysRev.177.1455,MEYER1979335,SENNHAUSER1981409,FRANKE1985351}, \cite{PhysRevC.33.2204,BELOZYOROV1986352,1987PZETF45196G,AmelinJEPT,PhysRevC.44.325,1995JETP,SIDORCHUK200454} and \cite{SIDORCHUK200454,2005EPJA24231G}, respectively.
The FY calculations \cite{Lazauskas2019} with  N$^3$LO, INOY and MT13 interactions are shown (right) for comparison.}\label{fig:1}
\end{figure*}

\section{Results and discussions}
The s.p.~Berggren basis is generated by a finite-depth Wood-Saxon (WS) potential including a spin-orbit coupling.
The parameters of the WS potential read $R_0 = 2.0$ fm for its radius, $a = 0.67$ fm for the diffuseness, $V_{ls} =  7.5$ MeV  for its spin-orbit strength, and $V_0 = -25$ MeV for the central depth.
With these parameters, the $0s_{1/2}$ and  $0p_{3/2}$ poles in the Berggren basis are bound and resonance states, respectively.
The single proton is well bound in $^4$H, so that we use the harmonic oscillator (HO) basis for proton, while the Berggren basis is used for neutron partial waves.
Alternatively, for the $^4$Li calculation, the HO basis is used for neutron, and the Berggren basis generates the proton part of the NCGSM model space.
Because proton and neutron emissions are both present in the excited states of $^4$He \cite{TILLEY19921}, the Berggren basis is used for both protons and neutrons therein.
As the continuum coupling induced by high partial waves is very weak, it can be neglected in NCGSM calculations.
Thus, we take a NCGSM model space generated by the $s_{1/2}$, $p_{3/2}$ and $p_{1/2}$ partial waves using the Berggren basis, while other partial waves, i.e., $d, f$ and $g$, are represented by HO basis states.
All HO one-body orbitals satisfy $2n+l \leq N_{\textrm{max}} = 14$.
The $L^+$ contour is defined by the coordinate points  $(0,0), (0.2, -0.15), (0.4, 0.0)$ and $(3.0,0.0)$ (all in fm$^{-1}$) for neutrons,
and $(0,0), (0.35, -0.15), (0.7, 0.0)$ and $(3.0,0.0)$ (all in fm$^{-1}$) for protons.
Each segment of the contour $L^+$ is discretized with 12 points, so that 36 discretization points are used in total for each partial wave.
NCGSM calculations are almost independent of the length parameter $b$ of the HO basis \cite{PhysRevC.88.044318,PhysRevLett.119.032501,PhysRevC.100.054313}, so that $b = 2$ fm is used  in this work.
We have checked that, for a given interaction, the g.s. energies of the well-bound $^3$H, $^3$He and $^4$He obtained with NCGSM are the same as those issued from NCSM calculations where the HO basis is used.

A complete diagonalization of the continuum NCGSM Hamiltonian is not possible because of the huge model space dimensions.
Consequently, we firstly calculated NCGSM eigenstates of $A = 4$ systems with the Berggren basis in a truncated space where three particles at most can occupy scattering states, to generate natural orbitals  \cite{NaturalOrbitals}.
As the natural orbitals are generated by the s.p.~density matrix associated with the targeted Hamiltonian eigenstates, they recapture a large part of the nuclear structure of many-body states \cite{NaturalOrbitals}.
One can then do NCGSM calculations without truncations using the basis of natural orbitals, as they are in a much smaller number than Berggren basis states. Each of natural orbitals is a combination of many different Berggren basis states.
It was checked in \cite{PhysRevC.100.054313} that the use of the complex-momentum natural orbitals provides virtually the same results as with the Berggren basis in NCGSM calculations. Therefore, natural orbitals offer an efficient basis to get calculations converged fast.

In our applications, the Argonne $v_{18}$ \cite{PhysRevC.51.38}, CD-Bonn \cite{PhysRevC.63.024001} and chiral N$^3$LO  \cite{PhysRevC.68.041001} potentials are renormalized using the SRG method with $\lambda = 2.0$ fm$^{-1}$,
providing  the softened interactions denoted by AV18-SRG2.0, CD-Bonn-SRG2.0 and N$^3$LO-SRG2.0, respectively.
We also employed the Daejeon16 potential \cite{SHIROKOV201687} which reproduces well various observables of light nuclei.
The Daejeon16 potential parameters are adjusted \cite{SHIROKOV201687} from the results arising from the use of the  N$^3$LO potential, which is renormalized by  SRG  using  $\lambda = 1.5$ fm$^{-1}$.
The latter renormalized N$^3$LO interaction, denoted by N$^3$LO-SRG1.5, is also used in the present calculations. 


The obtained energy and resonance width of the $^4$H ground state using NCGSM are presented in Fig.~\ref{fig:1},
along with experimental data \cite{COHEN1965242,PhysRev.177.1455,MEYER1979335,SENNHAUSER1981409,FRANKE1985351,PhysRevC.33.2204,BELOZYOROV1986352,1987PZETF45196G,AmelinJEPT,PhysRevC.44.325,1995JETP,SIDORCHUK200454,2005EPJA24231G} and the FY calculations \cite{Lazauskas2019}.
Several interactions are employed in the  NCGSM calculations (indicated in Fig.~\ref{fig:1}).
The calculations show that the N$^3$LO-SRG2.0, CD-Bonn-SRG2.0 and AV18-SRG2.0 interactions give similar results for the ground state of $^4$H, with an energy of about 1.7 MeV above the $^3$H$+n$ threshold and a width  close to 0.9 MeV.
Similar energies and widths are obtained when using N$^3$LO-SRG1.5 and  Daejeon16 potentials.
Compared to the above calculations with $\lambda=2.0$ fm$^{-1}$, however, the energies obtained with N$^3$LO-SRG1.5 and Daejeon16 are more bound by about 0.4 MeV and the widths are smaller by about 0.2 MeV.
Experimental data on the $^4$H resonance have been constantly emerging since 1960s \cite{COHEN1965242,PhysRev.177.1455,MEYER1979335,SENNHAUSER1981409,FRANKE1985351,PhysRevC.33.2204,BELOZYOROV1986352,1987PZETF45196G,AmelinJEPT,PhysRevC.44.325,1995JETP,SIDORCHUK200454,2005EPJA24231G}, but the data do not agree with each other. The experimental energy can vary from 1.6 MeV to 3.8 MeV, from one experiment to another, while the experimental width is even more uncertain  (see Fig.~\ref{fig:1}).
However, a recent experiment \cite{2005EPJA24231G} provides relatively small values for both energy and width with small experimental errors, which are close to our results.
The FY calculations with analytic continuation to the continuum \cite{Lazauskas2019} are also given in Fig.~\ref{fig:1}, showing a lower energy position and a broader width of the $^4$H resonance. In the analytic continuation method, the resonance observables cannot be determined directly, but are obtained with an extrapolation by artificially binding the system with an "external field" driven by an additional strength parameter \cite{Lazauskas2019}. The resonance width thus obtained  may depend on the extrapolation with the analytic continuation method  or on the angle with the complex scaling method. As the results obtained in NCGSM arise from a direct diagonalization of the complex Hamiltonian, represented by the Berggren basis, they neither depend on any additional parameter nor invoke any extrapolation approximation.

Both experimental data and our calculations suggest a first $1^-$ excited state which is very close to the $2^-$ ground state in energy for $^4$H.
The splitting of these two states is about 50 keV in our NCGSM calculations, and the obtained width of the $1^-$ excited state is about 1 MeV, which is comparable to that of the $^4$H ground state.
Our calculation indicates that both the ground state and first excited state of $^4$H  are mainly dominated by the unbound single neutron in the $p_{3/2}$ partial wave.
It seems to be quite probable that the experimental width of the $^4$H ground state appears like a mixture of the two resonances of the ground state and the first excited state.
In fact, the small splitting between the ground state and the first excited state in $^4$H raises many difficulties in experiment to distinguish  these two states.

\begin{figure}[!htb]
\includegraphics[width=1.0\columnwidth]{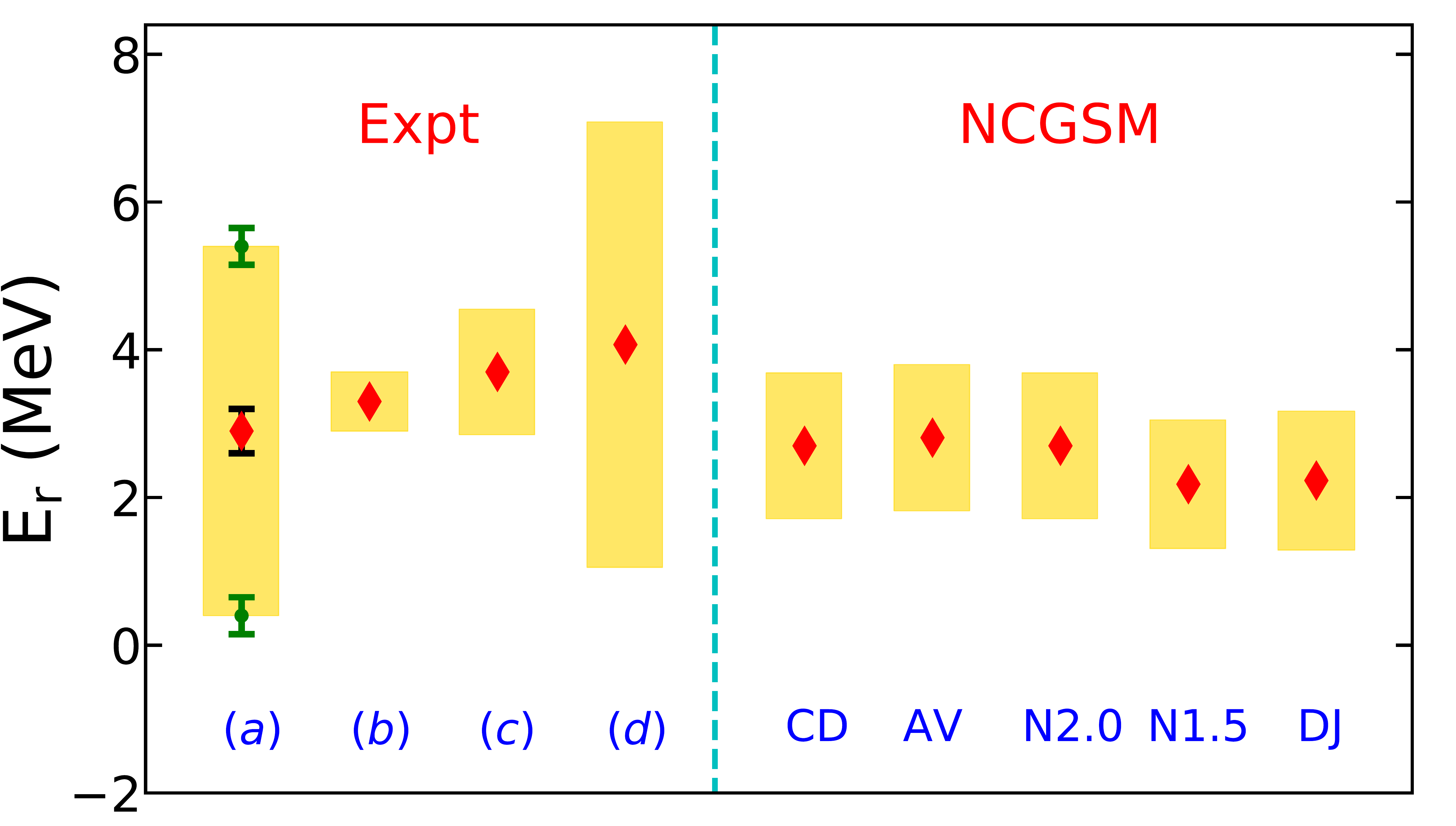}
\caption{Similar to Fig.1, but for $^4$Li, with experimental data: (a) \cite{PhysRevLett.15.300}, (b) \cite{PhysRevC.42.448}, (c) \cite{PhysRevC.42.550}, (d) \cite{TILLEY19921}.
Energy is given with respect to the $^3$He$+p$ threshold.}\label{fig:2}
\end{figure}

The  calculations of the $^4$Li ground state are shown in  Fig. \ref{fig:2}, along with experimental data \cite{PhysRevLett.15.300,PhysRevC.42.448,PhysRevC.42.550,TILLEY19921}.
The $^4$Li g.s. energy is about 2.7 MeV above the $^3$He+$p$ threshold, and the resonance width is about 2 MeV,  in the NCGSM calculations using  AV18-SRG2.0, CD-Bonn-SRG2.0 and N$^3$LO-SRG2.0.
The NCGSM calculations using N3LO-SRG1.5 and Daejeon16 provide an energy of about 2.2 MeV and a width of about 1.8 MeV.
Four different experiments \cite{PhysRevLett.15.300,PhysRevC.42.448,PhysRevC.42.550,TILLEY19921} gave different data on the $^4$Li energy, from 3 MeV to 4 MeV, and the data on width are much more uncertain, varying from 0.8 MeV to 6 MeV, from one experiment to another.
As seen in the mirror nucleus $^4$H, tSimilar to the situation in $^4$H, both the ground and $1^-$ excited states of $^4$Li have the $p_{3/2}$ character of the unbound single particle. The calculated width of the $1^-$ excited state is close to 2 MeV. Thus, it is also difficult to detect and distinguish experimentally the ground state and the first excited state in $^{4}$Li.

\begin{figure}[!htb]
\includegraphics[width=1.0\columnwidth]{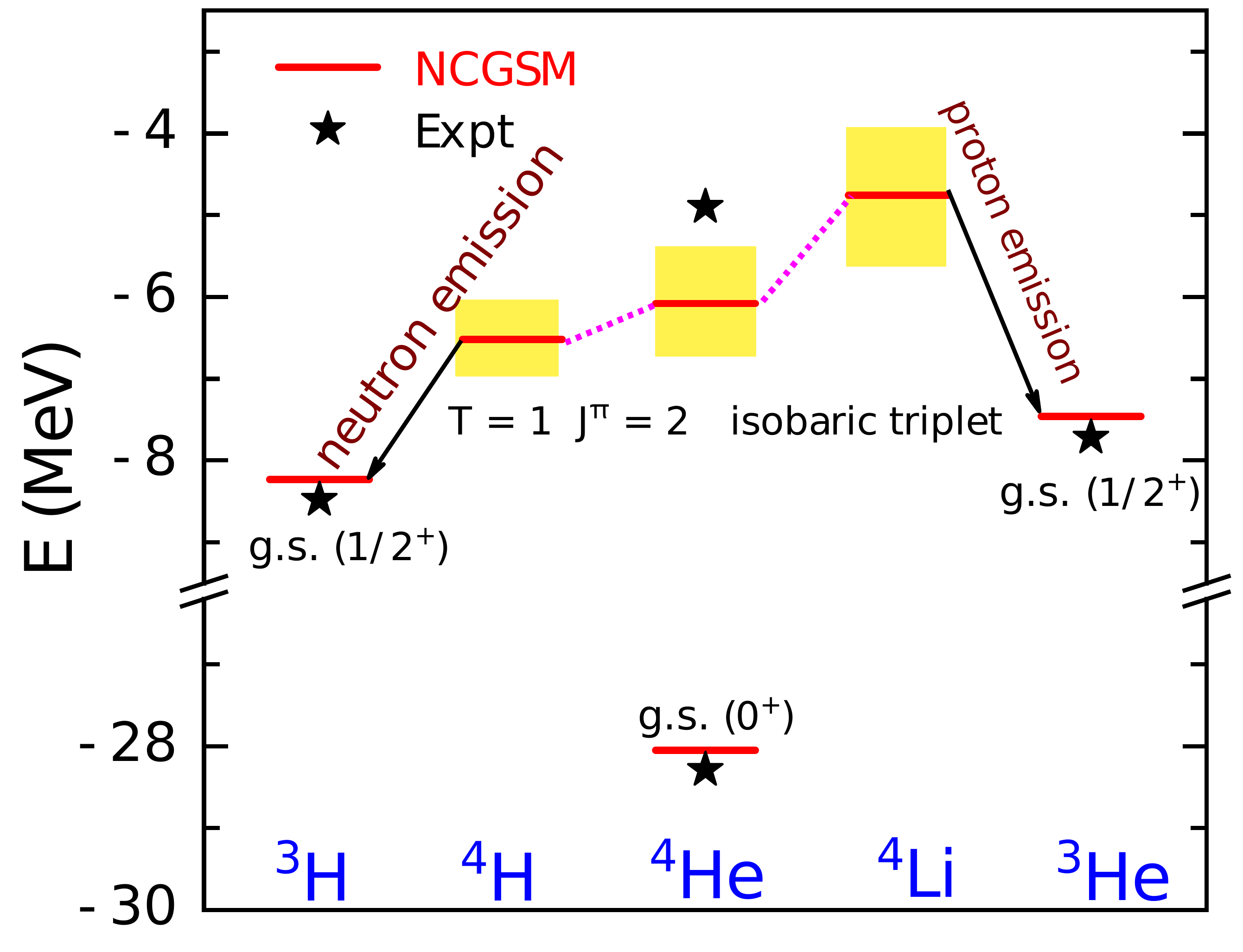}
\caption{Calculated level structures of $A$ = 3, 4 nuclei using NCGSM with the N$^3$LO-SRG2.0 interaction, and available experimental data \cite{TILLEY19921,ame}. The $T=1$ isobaric triplet states of $A= 4$ nuclei are connected by dotted lines.
  The experimental energy of the $T = 1$ $J^{\pi}$ = 2$^-$ state in $^4$He is deduced from the $R$-matrix analysis in \cite{TILLEY19921}.}\label{fig:3}
\end{figure}

The calculated ground states of $^4$H and $^4$Li are both unbound, with  large particle-emission widths.
Together with the first $T=1$ $J^\pi=2^-$ excited state of $^4$He, they form the $T=1 $ isospin triplet states of $A =4$  systems.
Figure  \ref{fig:3} presents the states in $A =3,4$ nuclei, with the isospin triplet states calculated with NCGSM using N$^3$LO-SRG2.0.
The results are also compared with experimental data \cite{TILLEY19921,ame}.
As the data on $^4$H and $^4$Li are not consistent with each other, they are not shown in Fig. \ref{fig:3}.
Our calculations provide good descriptions of the eigenstates of $A = 3,4$ nuclei.
The calculated energy of the $T=1$ $J^\pi=2^-$ excited state in $^4$He is close to the experimental datum extracted from an $R$-matrix analysis \cite{TILLEY19921}.
However, the decay width of this state, arising from the $R$-matrix analysis \cite{TILLEY19921}, is about three times larger than ours.
Our calculation gives the $^4$He  $2^-$ excited state at 22.03 MeV above the $^4$He ground state.
The result then agrees with the experimental datum in \cite{PhysRevLett.15.300}, which reported that this state is located at 22.5 $\pm$ 0.3 MeV.
The present calculations with N$^3$LO-SRG2.0 provide that the $T = 1$ isospin triplet states of $A =4$ nuclei are all unbound and bear broad decay widths,
which is consistent with the early experimental observations \cite{PhysRevLett.15.300}.

Isospin symmetry breaking of the $T = 1$ isospin triplet states in $A = 4$ nuclei, mainly caused by the repulsive Coulomb force, can be clearly seen in our calculations depicted in Fig. \ref{fig:3}.
The $2^-$ ground state of $^4$H, where the Coulomb force is absent, is the lowest state in the $T = 1$ isospin triplet.
$^4$H is more bound by about 0.5 and 1.7 MeV,  compared to the $2^-$ analog states in $^4$He and $^4$Li, respectively. 
The resonance width of the $2^-$ excited state in $^4$He is about 1.5 MeV, which lies between the widths of $^4$H and $^4$Li. Furthermore, to have an estimate of the isospin mixing in the $T=1$ triplet states, we have calculated the isospin expectation values of the states in the $A =4$ isobars. The results show that the isospins in $^4$Li and  $^4$H ground states are equal to 1, and that of the $^4$He ground state is equal to 0. However, the value is about 0.71 for the $2^-$ excited state of $^4$He,  which suggests that a large isospin mixing occurs in this state. The calculated isospin mixing in the $2^-$ excited state of $^4$He is consistent with the values obtained experimentally in \cite{PhysRevLett.57.543,PhysRevLett.72.3476}.
The $A=4$ isospin triplet states have been  investigated by the FY method \cite{PhysRevC.93.044004}.
In that approach, the FY equations are solved using a basis of $L^2$ states to calculate artificially bound four-body states, whose energy is then extrapolated to the negative separation energy region \cite{PhysRevC.93.044004}.
The extrapolated energies \cite{PhysRevC.93.044004} are close to those evaluated in the present NCGSM.
However, the method used in \cite{PhysRevC.93.044004} does not allow to calculate the resonance width.

In the present calculations, the three-nucleon force (3NF) is not included. It has been realized that the 3NF effect becomes more and more significant with increasing the mass number of nucleus. However, the inclusion of 3NF increases greatly the complexity of calculations, particularly in the many-body NCGSM calculation where the complex-energy plane is used with including resonance and continuum partial waves. In practical calculations, the 3NF effect depends on the form of the two-nucleon (2NF) interaction taken. For example, the nonlocal 2NF INOY potential without making an explicit use of 3NF can provide the excellent calculations of $A=3, 4$ binding energies in the framework of the FY equations \cite{PhysRevC.84.054010,PhysRevC.70.044002} and the $4N$ scatterings within the AGS equations \cite{PhysRevC.76.021001}. The off-shell behavior in nonlocal potentials seems to play a role similar to 3NF. It was shown that the nonlocal N$^3$LO interaction \cite{PhysRevC.68.041001} without invoking 3NF can well describe the $n$-$^3$H scatterings near the resonance peak \cite{PhysRevC.84.054010}. In \cite{PhysRevC.93.044004}, the authors, using the FY method with the Argonne potential, have investigated the $A=4$ nuclear systems, finding that the $T=3/2$ 3NF effect is very weak in $^4$H, $^4$He and $^4$Li. Nevertheless, the nonlocality in the CD-Bonn or chiral EFT potential cannot fully replace the role of 3NF, though the calculations can be considerably improved with the nonlocal interactions \cite{PhysRevLett.85.944,PhysRevC.70.044002}. However, nonlocal potentials require relatively weak 3NF in calculations \cite{PhysRevLett.85.944,PhysRevC.70.044002}. From the present calculations, all the interactions used can provide good descriptions of the $A = 3, 4$ systems. Therefore, we may assume that the 3NF effect can be relatively small in the $A=4$ resonance states which have $n$($p$)-$^3$H($^3$He) resonance structures. However, the 3NF effect in the tightly bound $^4$He ground state can be noticeably larger, and increases the binding energy of the ground state, indicating a lower position of the calculated $^4$He ground state in the level scheme shown in Fig. \ref{fig:3}. The calculated energy of the $T=1$ $J^\pi=2^-$ isobaric analog in $^4$He is about 1 MeV lower than the datum \cite{TILLEY19921}. However, as discussed, the calculation of the excitation energy of the isobaric analog can be improved by a lower ground-state energy. The excitation energy plays a more direct role in describing excited states. Added to that, the experimental energy of the isobaric analog in $^4$He was extracted using the $R$-matrix approximation \cite{TILLEY19921}, and hence a large experimental uncertainty may be expected. As a whole, the exclusion of 3NF should not change the discussions of physics problems related to the $A=4$ resonances.

\section{Summary} 
$A=4$ isobars are challenging from both theoretical and experimental points of view.
Due to their small number of nucleons, $A=4$ nuclear states can be calculated with \textit{ab initio} NCGSM, so that their structure can be precisely evaluated.
The internucleon correlations and continuum coupling are both taken into account in NCGSM calculations.
However, experimental data involving the $A=4$ isobars do not agree with each other, especially for the resonance width which can vary by several MeV from one experiment to another.

Consequently, \textit{ab~initio} NCGSM calculations of unbound $A=4$ nuclear states have been done in order to theoretically clarify  the situation.
For this, the unbound $2^-$ ground states of $^4$H and $^4$Li, as well as the $T=1$ $J^\pi = 2^-$ isobaric analog state in $^4$He, have been calculated with NCGSM using different modern realistic forces.
Due to their proximity to the $2^-$ ground states, the unbound $1^-$ excited states of $^4$H and $^4$Li have been calculated as well.
Our calculations tend to favor the small widths provided by experimental data.
The first excited state in  $^4$H and $^4$Li is a broad $1^-$ state in our calculations, whose energy is about 50 and 100 keV above that of the $^4$H and $^4$Li ground states, respectively, and whose width is close to 1 and 2 MeV, respectively.
This is also in accordance with experimental observations.
The $T = 1$ isospin triplet  in $A=4$ nuclei has been studied. It consists of the $2^-$ground states of $^4$H and $^4$Li and of the $2^-$ excited state of $^4$He.
We have shown that the $2^-$ excited state of $^4$He is located in energy in the middle of the $2^-$ ground states of $^4$H and $^4$Li, and its width is also between those of $^4$H and $^4$Li.
Consequently, our calculations clearly exhibit isospin symmetry breaking in the $T = 1$ isospin triplet of the $2^-$ states in $A=4$ nuclei, whose origin can be attributed to both the presences of the Coulomb interaction and continuum coupling.
Their effects are, however, difficult to disentangle. 
The present \textit{ab initio} calculations provide promising information for further experimental studies of the unbound resonance systems of $A=4$ $T=1$ isobaric nuclei.

\begin{acknowledgments}
We thank J. P. Vary and P. Maris for providing us the Daejeon16 interaction.
This work has been supported by the National Key R\&D Program of China under Grant No. 2018YFA0404401; the National Natural Science Foundation of China under Grants No. 11835001,  11921006, 12035001, and  11975282; the State Key Laboratory of Nuclear Physics and Technology, Peking University under Grant No. NPT2020ZZ01 and NPT2020KFY13; the Strategic Priority Research Program of Chinese Academy of Sciences under Grant No. XDB34000000; 
the Key Research Program of the Chinese Academy of Sciences under Grant No. XDPB15;
and the CUSTIPEN (China-U.S. Theory Institute for Physics with Exotic Nuclei) funded by the U.S. Department of Energy, Office of Science under Grant No. de-sc0009971. The High-Performance Computing Platform of Peking University is acknowledged.

\end{acknowledgments}

\bibliography{NCGSM_A4}

\end{document}